\documentclass[twocolumn,showpacs,floatfix,prl,superscriptaddress]{revtex4}

\usepackage{float}
\usepackage{graphicx}
\usepackage{bm}
\usepackage{amsmath}
\usepackage{amssymb}

\newcommand{\ts}{{\tilde \sigma}}
\newcommand{\beq}{\begin{eqnarray}}
\newcommand{\eeq}{\end{eqnarray}}
\newcommand{\beqq}{\begin{eqnarray*}}
\newcommand{\eeqq}{\end{eqnarray*}}

\begin{document}

\title{Bulk Topological Proximity Effect}


\author{Timothy H. Hsieh}
\email{thsieh@kitp.ucsb.edu}
\affiliation{
Kavli Institute for Theoretical Physics, University of California, Santa Barbara, California 93106, USA}
\author{Hiroaki Ishizuka}
\affiliation{
Kavli Institute for Theoretical Physics, University of California, Santa Barbara, California 93106, USA}
\author{Leon Balents}
\affiliation{
Kavli Institute for Theoretical Physics, University of California, Santa Barbara, California 93106, USA}
\author{Taylor L. Hughes}
\affiliation{
Department of Physics, Institute for Condensed Matter Theory,
University of Illinois at Urbana-Champaign, IL 61801, USA
} 

\begin{abstract}

Existing proximity effects stem from systems with a local order parameter, such as a local magnetic moment or a local superconducting pairing amplitude.  Here, we demonstrate that despite lacking a local order parameter, topological phases also may give rise to a proximity effect of a distinctively inverted nature.  We focus on a general construction in which a topological phase is extensively coupled to a second system, and we argue that in many cases, the inverse topological order will be induced on the second system. 
To support our arguments, we rigorously establish this ``bulk topological proximity effect'' for all gapped free fermion topological phases and representative integrable models of interacting topological phases.  We present a terrace construction which illustrates the phenomenological consequences of this proximity effect.  Finally, we discuss generalizations beyond our framework, including how intrinsic topological order may also exhibit this effect.    

\end{abstract}

\maketitle

Topological phases of matter cannot be characterized by a local order parameter, unlike conventional ordered phases with, e.g., magnetic order.  Instead, one discriminates them from so-called trivial phases of matter by analyzing either their global properties, such as topological invariants \cite{tknn, hasankane, qizhang, moore, niuwen} and entanglement fingerprints\cite{hamma1,hamma2, levinwen, kitaev,li2008,bernevig2009a,bernevig2009,flammia2009,pollmann2010,fidkowski2010,turner2010,hughes2011,bernevig2011,aris2011,zhang}, or their boundaries, which often exhibit robust gapless states.  Despite such subtleties in characterization and classification, significant progress has been made in understanding the different varieties and phenomenology of topological phases.  One broad class of states are those with intrinsic topological order \cite{wen}, which possess excitations with fractional statistics, topological ground state degeneracy, and long range entanglement.  Additionally, even systems without intrinsic topological order can support robust topological phases in the presence of a symmetry.  In particular, two gapped phases which cannot be smoothly connected without either closing the gap, or breaking a given symmetry, constitute distinct symmetry protected topological (SPT) phases \cite{guwen, pollmann, chen}.  Both classes of topological matter--intrinsic and SPT--have been realized respectively in fractional quantum Hall systems \cite{tsui} and more recently in a variety of topological insulators (TIs) \cite{hasankane}.  

Given the maturity of material synthesis and engineering processes, one can easily imagine fabricating topological materials in proximity to other systems. Indeed, depositing superconductors and magnets on 2D and 3D time-reversal invariant topological insulators has become a booming industry\cite{fukane,aliceareview,veldhorst,williams2012}, as the superconducting proximity effect has featured prominently in proposals for realizing Majorana fermions in solid-state materials\cite{fukane}.  However, the opposite effect--how the topological state might in turn affect proximate systems-- has not been explored to the same extent.  In conventional proximity effects, such as those involving superconducting or magnetic phases, there is an order parameter which penetrates into a proximate material. Is this phenomenon different, or denied, for topological phases which do not carry a local order parameter?   
     

In this work, we demonstrate that, despite lacking a local order parameter, topological phases can nonetheless exhibit a proximity effect, in which a topologically nontrivial system causes a proximate system to become topologically nontrivial as well.  In many cases, the induced topological phase of the proximate system will be an ``inverse'' of the original phase, to be made precise shortly.  In other cases, the {\it entire} combined system can be driven into a new topological phase by simply increasing the inter-system coupling.  To avoid confusion, we emphasize that this phenomenon is fundamentally different from the ``topological proximity effect'' discussed in \cite{tpe}, where the gapless boundary states of a three dimensional topological insulator essentially move into a proximate thin metallic film.  In contrast, the setup we envision involves coupling two {\it bulk} systems {\it of the same dimension} to each other, and inducing a full bulk topological phase; it is not just an effect on the boundary states.  We will hereafter refer to this phenomenon as the bulk topological proximity effect (BTPE).  

The structure of this Letter is as follows.  We first detail the setup for the BTPE and subsequently present general arguments for why it is expected to work for a large class of topological phases and inter-system couplings.  For concreteness, we then demonstrate this phenomenon in a bi-layer of free fermion systems in which one system with Chern number $-1$ induces Chern number $+1$ in the second system.  After rigorously generalizing to all free fermion topological phases, we next argue how this effect may be realized in coupled spin chains, and explicitly demonstrate it in a one-dimensional SPT protected by $Z_2 \times Z_2$ symmetry.  Finally, we discuss generalizations of the BTPE, including applications to intrinsic topological order.        

{\bf General framework:} Our setup is a composite system consisting of subsystems $A$ and $B$, with identical Hilbert spaces.  However, we will be interested in Hamiltonians for the combined system
\beq
H = H_A + H_B +H_{AB}
\eeq
in which $H_A$ and $H_B$ are very different: $H_A$ will be a gapped Hamiltonian with a topologically nontrivial ground state and an energy gap of order $\Delta_A$, while $H_B$ is \emph{any} Hamiltonian with a much smaller energy scale $W_B\ll\Delta_A.$  $H_{AB}$ is a coupling between the subsystems with strength $g_{AB}$ which we assume obeys $W_B\ll g_{AB}\ll\Delta_A$.  The motivation for keeping $g_{AB}\ll\Delta_A$ is that we would like to have a notion of $A$ and $B$ as two independent subsystems (meaning the joint ground state approximately factorizes: $|\psi^0_{AB}\rangle \approx |\psi^0_A\rangle|\psi^0_B\rangle$), and we require $W_B\ll g_{AB}$ so that subsystem $B$ is ``susceptible" to $A$'s topological Hamiltonian. The precise nature of $H_B$ is irrelevant in the strong inter-system coupling limit, and for simplicity we will focus on the extreme limit $H_B=0$; in other words, without the coupling, system $B$ is a set of independent degrees of freedom, as in a Kondo lattice model.  Having a non-vanishing $H_B$ can lead to richer phase diagrams in the intermediate coupling regime, and we also give an example of this below.

\begin{figure}
\centering
\includegraphics[height=2in]{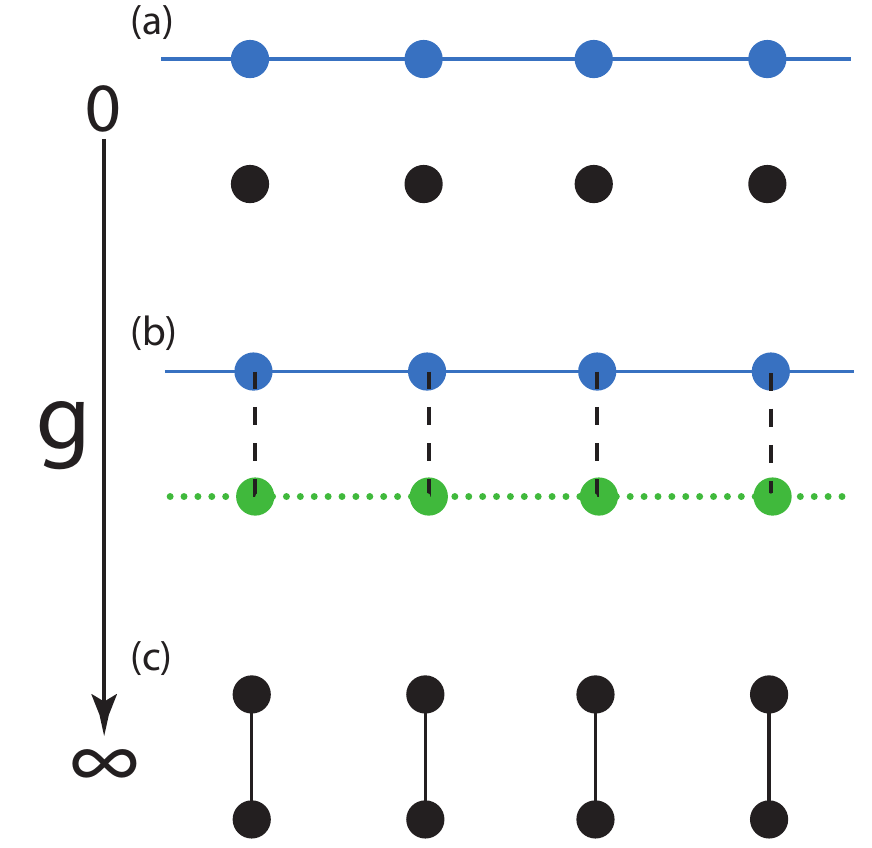}
\caption{Schematic of the bulk topological proximity effect.  (a) The starting point is a topological state in system $A$ (blue) and free, isolated degrees of freedom in system $B$ (black). (b) At small coupling between $A$ and $B$ (dashed lines), the ``inverse'' topological phase is induced in system $B$, and the composite is trivial.  (c) At infinite coupling, the system is a product state. }
\end{figure}


Let us first focus, for the sake of argument, on a particular (and large) class of topological Hamiltonians and a coupling $H_{AB}$ that could be dominant in realistic systems. Specifically, we consider any Hamiltonian $H_A$ which has an ``invertible'' topological ground state $\psi_A$, i.e., there must exist an ``inverse'' $\overline{\psi_A}$ such that the direct product of the two $ \psi_A \otimes \overline{\psi_A}$ is a topologically trivial state \cite{inverse1, inverse2, inverse3}.  Examples of invertible states are all SPTs, including fermionic SPTs such as Chern insulators, and Kitaev/Majorana wires.  As for the coupling $H_{AB}$, we focus on those whose ground state is a product state of maximally entangled $A$ and $B$ sites, as depicted in Fig. 1c.  Such couplings are prevalent and include, e.g.,  inter-layer tunneling in bilayer fermion systems and antiferromagnetic exchange coupling for bilayer spin systems. 

With these mild assumptions, at infinite coupling between $A$ and $B$, the combined system is topologically trivial because it is a product state.  As the coupling is decreased, naively, there could be a phase transition to a topologically non-trivial ground state. In this paper, however, we demonstrate that the strong and weak coupling phases are smoothly connected for \emph{all} free fermion systems and a class of interacting integrable models below. Since this implies that the weak coupling phase of the composite system is also topologically trivial, system $B$ must carry the inverse topological order as system $A$, even at weak coupling, thus realizing BTPE. From another perspective, system $B$ `screens' the topological phase of system $A$ so that the composite system is neutralized/trivial.  The above holds for $H_B=0$; if $H_B\neq 0$, system $B$ can potentially over-screen $A$ and induce new topological properties for the composite system.


{\bf Free fermion topological proximity effect:} We now provide a concrete example of the topological proximity effect due to a Chern insulator.  Specifically, we use the following tight-binding model for spin-1/2 fermions on the square lattice:  
\beq
H_{CI}(\mu) &=& \sum_k h_{\alpha\beta} (k) c^{\dagger}_{Ak\alpha} c^{\vphantom\dagger}_{Ak\beta} \label{ham} \\
h(k) &=& (\cos{k_x} + \cos{k_y} - \mu)\sigma^z + \sin{k_x} \sigma^x +\sin{k_y} \sigma^y.  \nonumber
\eeq
Here $c^{\vphantom\dagger}_{Ak\sigma}$ ($c_{Ak\sigma}^\dagger$) is the fermion annihilation (creation) operator with wave number $k$ and spin $\sigma$ on layer $A$ and $\sigma^{x,y,z}$ are the Pauli spin matrices.  For $0<\mu<2$, the ground state has Chern number $-1$ and for $\mu>2$, the ground state is trivial.  
The full Hamiltonian is given by:
\beq
H_A&=& H_{CI}(\mu=1), \;\;\; H_B=0 \nonumber \\ 
H_{AB} &=& g \sum_i c^{\dagger}_{Ai\sigma} c^{\vphantom\dagger}_{Bi\sigma} + h.c. \nonumber
\eeq

For weak coupling, degenerate perturbation theory to second order provides an effective Hamiltonian for $B$ given by 
\beq
H^{\rm eff}_B &=& \sum_k h^{\rm eff}_{\alpha\beta} (k) c^{\dagger}_{Bk\alpha} c^{\vphantom\dagger}_{Bk\beta}, \\
h^{\rm eff}(k) &=& -\frac{4g^2}{\Delta(k)} |\psi^{\rm ex}_k\rangle \langle \psi^{\rm ex}_k|,
\eeq 
where $|\psi^{\rm ex}_k\rangle$ is the spinor of the excited state of $h(k),$ and $\Delta(k)$ is its energy difference from the ground state of $h(k)$.  However, the conduction band of $H_A$ must have the opposite Chern number of the valence band (if both bands are filled, then system $A$ would be a trivial insulator). Thus, we find that the effective ground state of $H_B$ has Chern number $+1$.

\begin{figure}
\centering
\includegraphics[height=2.5in]{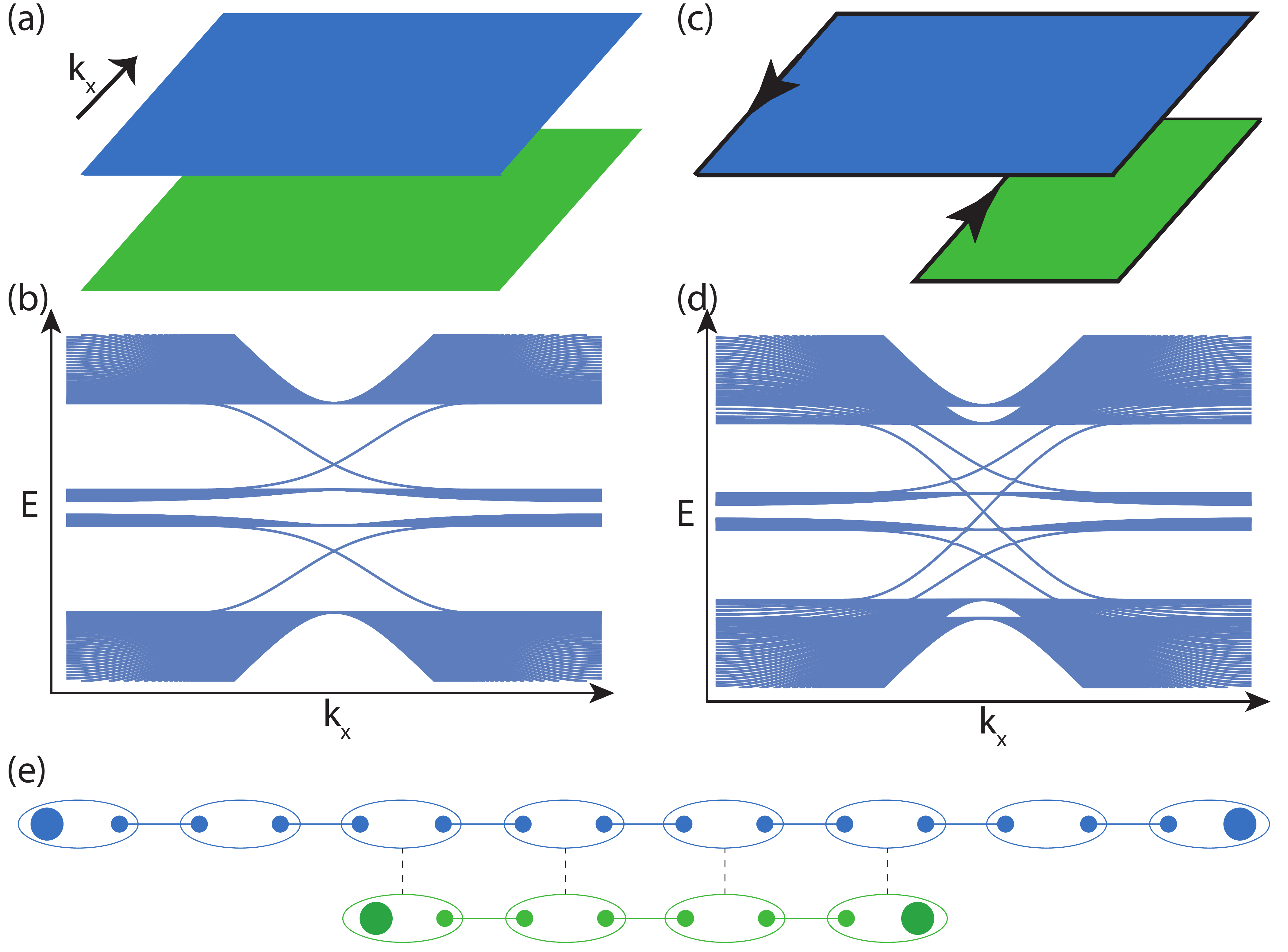}
\caption{Topological proximity effect of a Chern insulator. (a) A bilayer slab (in)finite in the $(x)y$ direction, with spectrum shown in (b). (c) A ``terrace'' in which the left boundary of $A$ is separated from that of $B$.  (d) The resulting spectrum, with weakly coupled counter-propagating edge states in the bulk gap localized at the distinct boundaries (bold arrows in (c)). (e) Schematic of a Kitaev/Majorana wire inducing an inverse phase on a proximate chain.  In addition to the original Majorana zero modes (large blue dots), there will new Majorana modes (large green dots). }
\end{figure}
    
In fact, in all free fermion models of the above type, in which a nontrivial topological system $A$ is tunnel coupled to a flat band in the band gap of system $A$, the strong and weak coupling limits are smoothly connected without any phase transition.  This is because the full single-particle Hamiltonian has the form
\beq
H = \left( \begin{array}{cc}
H_A & g \\
g & 0
\end{array} \right)=\tfrac{1}{2}(\mathbb{I}+\tau^z)\otimes H_A+g(\tau^x\otimes\mathbb{I}),
\eeq 
where the upper-left and lower-right blocks correspond to the local basis of systems $A$ and $B$, respectively (for convenience we have centered the band gap of $A$ and the flat band(s) of B at zero energy.).  However, even after $H_A$ is diagonalized ($H^d_A = U H_A U^{-1}$), the form of the above matrix can be preserved by changing the basis of $B$ with the same unitary $U$: 
\beq
H = {\sf U}^{-1}
\left( \begin{array}{cc}
H^d_A & g \\
g & 0
\end{array} \right)
{\sf U},
\eeq
with ${\sf U} = \mathbb{I}\otimes U$.
The resulting eigenvalues are $E_n^\pm=(E^0_n/2) \pm \sqrt{(E^0_n/2)^2 + g^2}$, where $E_n^0$ are the eigenvalues of $H_A$.  Hence, all eigenvalues are repelled from $E=0$ if $g\ne0$, implying that there is no phase transition between strong and weak coupling. This unambiguously establishes the BTPE for all free fermion topological phases.

The interpretation of topological ``screening'' can be explicitly shown from the ground state wave function
\beq
\left|\psi\right>&=&\prod_{E_n<0} \left( \cos\frac{\theta_n}2\left|\varphi^{\rm fill}_n\right>_A + \sin\frac{\theta_n}2\left|\varphi^{\rm fill}_n\right>_B \right)\nonumber\\
&&\times\prod_{E_n>0} \left( \sin\frac{\theta_n}2\left|\varphi^{\rm ex}_n\right>_A - \cos\frac{\theta_n}2\left|\varphi^{\rm ex}_n\right>_B \right),
\eeq
where $\tan\theta_n=g/E_n^0$ and $\left|\varphi^\alpha_n\right>_a$ ($\alpha=$ex, fill, $a=A,B$) is the single-particle wave function for excited (ex) and filled (fill) states of $H_A$ at $g=0$; the index $a$ denotes the layer the electron is on. As $g\to0$ ($\theta_n\to0$), the ground state wave function is approximately the direct product of a $C=-1$ band of electrons in $A$ and a $C=1$ band in $B$, consistent with the perturbation argument above. We note that, for arbitrary $g$, the Berry curvature for the bands in the first and second products are exactly the same as that of filled and excited bands of $H_A$ at $g=0$, respectively. Hence, for arbitrary $g$, the Chern number for the entire system is zero.

Since the composite system is trivial, the nontrivial nature of $B$ is not immediately manifest.  If $A$ and $B$ are strictly identical, then the composite system with boundary will have a gapped edge, since the gapless mode of $A$ hybridizes with the counter-propagating mode of $B$ (see Fig. 2a,b).  However, in a ``terrace'' construction in which the boundaries of $A$ and $B$ are separated, as shown in Fig. 2c, the gapless modes on the two different boundaries interact weakly and are exhibited clearly in Fig. 2c,d.  We note that in this terrace construction, increasing the coupling shifts the weight of the edge mode initially on the $B$ boundary onto the $A$ side of the kink.  This construction will work for any system, and we provide a schematic illustration for a proximity effect due to a Kitaev chain in Fig. 2e, in which Majorana bound states are induced at the ends of system $B$ when using a terrace construction. 

In realistic models $H_B$ is likely to be non-zero, but  as long as the inter-system coupling $g$ is much larger than the energy scale(s) in $H_B$, then the BTPE of the above type will occur.  However, the properties of the small-$g$ regime depend on the nature of $H_B,$ and may exhibit a wide range of phases.  To illustrate this in the Chern insulator example, we also considered $H_B =\gamma H_{CI}(\mu)$ of the form (\ref{ham}), but scaled down by a factor $\gamma$ ($0<\gamma<1$) relative to $H_A$.  The effect of coupling systems $A$ and $B$ is shown in Fig. 3a; depending on the coupling $g,$ and the intrinsic phase of $B$, parameterized by $\mu$, a variety of Chern numbers for the composite system can be achieved.  It is striking that at intermediate coupling, the {\it composite} system can carry the {\it opposite} Chern number of the original topological phase.  In this case, for a slab of the two systems, the gapless edge mode remarkably reverses direction at finite coupling (Fig. 3b). This effect can be thought of as an over-screening of the topological phase of $H_A,$ where instead of canceling the Chern number it induces a composite Chern number of the opposite sign. Optimistically, one could design a device utilizing this effect: since the chirality is controlled by the inter-layer coupling, one could imagine creating a low-dissipation, pressure-switchable diode, where the pressure modifies the inter-layer tunneling to switch the easy-current direction. 

\begin{figure}
\centering
\includegraphics[height=2.2in]{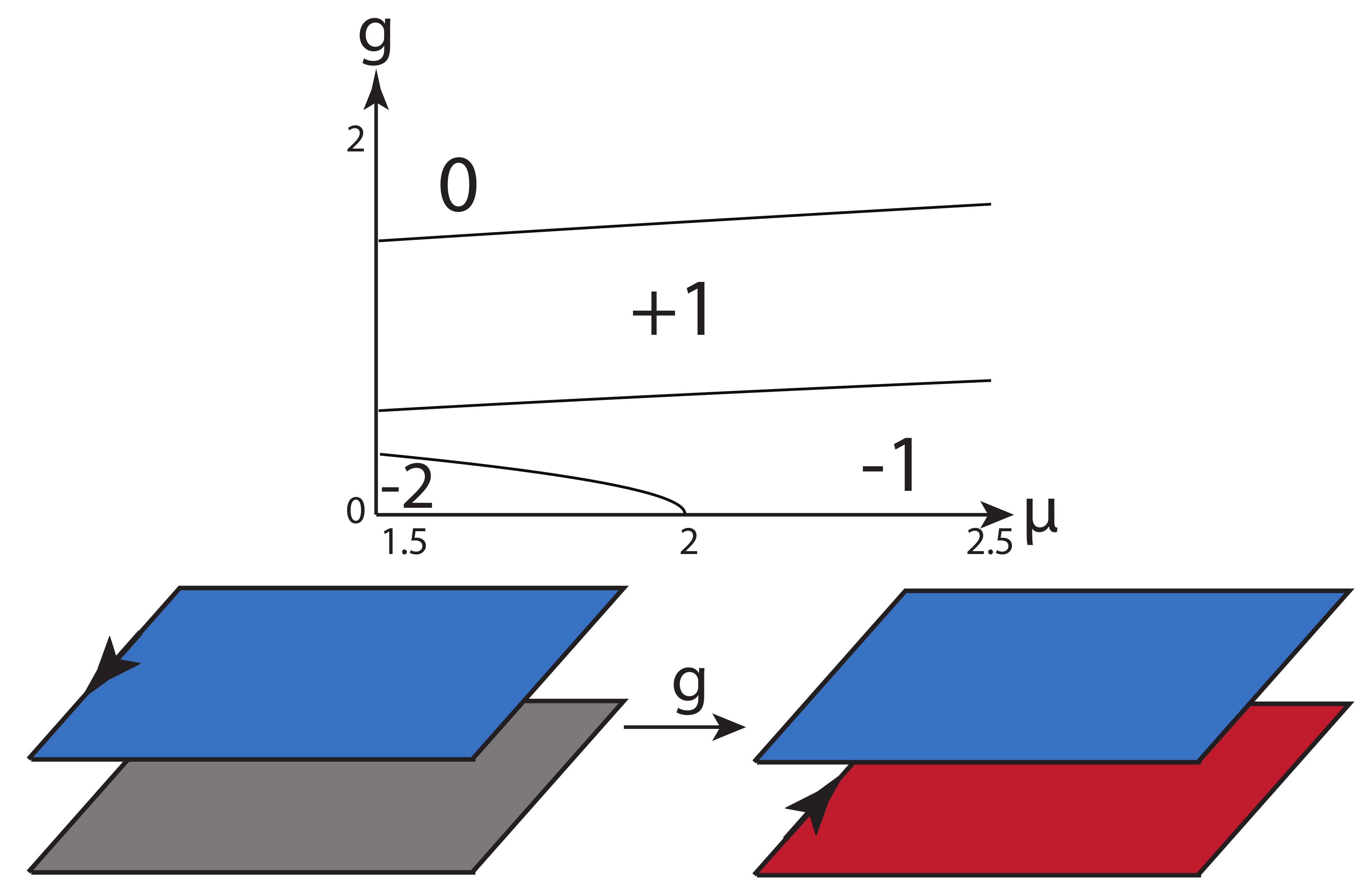}
\caption{Effect of nonzero Hamiltonian for system $B$.  (a) Taking $H_A=H_{CI}(\mu=1)$ (Chern number $-1$ ground state) and $H_B = 0.2H_{CI}(\mu)$, we calculated the phase diagram as a function of coupling $g$ and $\mu$.  The phases are labeled by the total Chern number of the composite system. (b) A cartoon illustrating how intermediate coupling can completely reverse the Chern number and hence the direction of the edge state.}
\end{figure}

{\bf Interacting topological proximity effect:} We expect the BTPE to occur generally in interacting systems as well. To illustrate, let us begin with a concrete model in which the effect can be explicitly demonstrated.  For system $A$, we choose the ``cluster state'' \cite{cluster} or ``$ZXZ$'' Hamiltonian of a spin-1/2 chain: 
\beq
H_A = -\sum_i \sigma^z_i \sigma^x_{i+1} \sigma^z_{i+2} .
\eeq
This system possess a $Z_2 \times Z_2$ symmetry generated by independent twofold rotations on the two alternating sublattices $a$ and $b$: $g_1 = \prod_{j\in a} e^{{\rm i} \pi \sigma^x_j/2}$, $g_2 = \prod_{j\in b} e^{{\rm i} \pi \sigma^x_j/2}$.  One dimensional systems with this symmetry have a $Z_2$ topological classification, and the above Hamiltonian provides an example of the nontrivial phase.  
Now we add another spin-1/2 chain $B$, with $H_B=0$, and introduce the coupling
\beq
H_{AB} = g \sum_i \left(\sigma^z_i \ts^z_i + \sigma^x_i \ts^x_i\right), \label{coupling}
\eeq
where the operators with tildes act on the $B$ spins.  Per our framework, such a coupling, when dominant, locks together the corresponding $A$ and $B$ spins into singlets. 

Due to the fact that $H_A$ is integrable (all eigenstates can be labeled by the conserved quantities $\sigma^z_i \sigma^x_{i+1} \sigma^z_{i+2} = \pm 1$), degenerate perturbation theory can be performed exactly.  Since all one-body and two-body operators anticommute with at least one conserved quantity, the lowest order effect occurs at $O(g^3)$ and yields the effective Hamiltonian for $B$:
\beq
H^{eff}_B= g^3 \sum_i \ts^z_i \ts^x_{i+1} \ts^z_{i+2}.
\eeq
Hence, system $B$ precisely inherits the topological order of system $A$ (a $Z_2$ SPT is its own inverse).

This behavior generalizes to other interacting SPTs, such as a spin-1 chain in the Haldane phase \cite{haldane} coupled to a Kondo lattice chain of isolated spin-1 sites via antiferromagnetic exchange.  In this case, second order perturbation theory will imprint the exponentially decaying correlations of chain $A$ onto the effective Hamiltonian of chain $B$, and we expect that the resulting effective nearest neighbor antiferromagnetic interaction will place chain $B$ in the Haldane phase as well.    

{\bf Beyond the framework:}  
While the assumptions specified and exemplified above are useful for understanding a large class of topological phases and their couplings to proximate systems, it is interesting to consider relaxing those assumptions.  In particular, thus far we have considered identical Hilbert spaces for systems $A$ and $B$, but allowing $B$ to be a different lattice, dimension, or even particle type, may give rise to new phenomena when coupled to $A$.  In a similar vein, the current framework specializes to couplings which, when infinite, maximally entangle corresponding degrees of freedom of $A$ and $B$, and the myriad ways of relaxing this constraint, especially when $A$ and $B$ are different Hilbert spaces, may prove fruitful.  For example one may imagine coupling a bosonic SPT state (system $A$) to a fermionic auxiliary system (system $B$) or vice-versa \cite{progress}. We generally expect that when the degrees of freedom of system $B$ are capable of screening the topological nature of system A then we will see similar proximity effects. 

Another interesting direction to consider are invertible topological phases with non-unique inverses. For example, system A could be \emph{stably}-equivalent to a conventionally inequivalent state, i.e., only equivalent after being combined with additional trivial states. These systems might induce a ``stable" proximity effect where they are screened not by their conventional inverse, but by a stably-equivalent inverse.  For example, a $\nu=8$ fermionic integer quantum Hall state could be screened by a free-fermion $\nu=-8$ state, or the (anti-chiral) bosonic integer quantum Hall $E_8$ state to which its inverse is stably equivalent\cite{mulligan2013,cano2014}. We expect that the choice of coupling terms between A and B would determine this outcome. Since there are many examples of stable equivalence, we expect them to give rise to a rich set of phase diagrams. 

Finally, we note that systems with intrinsic topological order may also exhibit the BTPE, albeit of a different variety than that considered above.  For example, consider two-dimensional $Z_2$ topological order, realized by the ``toric code'' Hamiltonian \cite{toric} for spin-1/2s on the links of a square lattice:
\beq
H_A = -\sum_s \prod_{l\in s} \sigma^z_l - \sum_p \prod_{l\in p} \sigma^x_l,
\eeq
where $s$ denotes stars of four links emerging from each vertex, and $p$ denotes spins on the square plaquettes.  Like the cluster state model, all the terms in the Hamiltonian commute, and thus provide conserved quantities to label eigenstates.  However, given periodic boundary conditions, there are an additional two global conserved quantities which endow $A$ with a fourfold ground state degeneracy.

Upon adding an identical Hilbert space $B$ of free spins $\ts$ and coupling $A$ and $B$ with the Hamiltonian in (\ref{coupling}), the lowest order effect [at $O(g^4)$] is to induce an effective Hamiltonian in $B$ given by the same star and plaquette terms above.  Hence, at small coupling, there is a proximity effect in which the ground state of the composite system is two copies of the toric code ground states.  In contrast to the above framework for invertible topological order, these two copies are not smoothly connected to the trivial product state of singlets found at infinite coupling -- for example, the ground state degeneracy at weak coupling is 16-fold.  As a result, there must be at least one topological phase transition at intermediate $g$.   How this occurs is non-trivial and would be an interesting subject for future study.
    
\acknowledgements{We thank Yuan-Ming Lu, Andreas Ludwig, and Chetan Nayak for interesting discussions.  THH was supported by a fellowship from the Gordon and Betty Moore Foundation (Grant 4304) and acknowledges a KITP graduate fellowship and DOE Office of Basic Energy Sciences Award No. DE-SC0010526. H.I. was supported by the JSPS Postdoctoral Fellowship for Research Abroad and by the MRSEC Program of the National Science Foundation under Award No.DMR-1121053. L.B. was supported by the National Science Foundation under grant No.DMR-12-06809. TLH thanks the hospitality of the KITP at UCSB where part of this work was completed during the Entanglement and Strongly Correlated Matter program, and thanks NSF CAREER DMR-1351895 for support.}

\end{document}